\documentclass{aa}  
\usepackage{natbib}
\bibpunct{(}{)}{;}{a}{}{,} 
\usepackage{graphicx}
\usepackage{txfonts}
\usepackage{color}
\newcommand{\sag}{\textsc{\large{sag}}}
\newcommand{\sagmdpl}{\textsc{\large{sag-mdpl}2}}
\newcommand{\roger}{\textsc{\large{roger}}}
\newcommand{\pyroger}{\textsc{\large{pyroger}}}

\begin{document}

   \title{Improving the accuracy of observable distributions for galaxies classified in the projected phase space 
   diagram}
   \titlerunning{Improving distributions of galaxy observables}


   \author{
            H\'ector J. Mart\'inez\thanks{hjmartinez@unc.edu.ar}\inst{\ref{iate},\ref{oac}}
            \and
            Mart\'in de los Rios\inst{\ref{uam},\ref{ift},\ref{sissa}}
            \and
            Valeria Coenda\inst{\ref{iate},\ref{oac}}
            \and
            Hern\'an Muriel\inst{\ref{iate},\ref{oac}}
            \and
            Andr\'es N. Ruiz\inst{\ref{iate},\ref{oac}}
            \and
            Sof\'ia A. Cora\inst{\ref{ialp},\ref{fcaglp}}
            \and
            Cristian A. Vega-Mart\'inez\inst{\ref{ucc}}
        }
\institute{
Instituto de Astronom\'ia Te\'orica y Experimental (CONICET -- UNC), 
Laprida 854, X5000BGR, C\'ordoba, Argentina\label{iate}
\and
Observatorio Astron\'omico, Universidad Nacional de C\'ordoba,
Laprida 854, X5000BGR, C\'ordoba, Argentina\label{oac}
\and
Departamento de F\'isica Te\'orica, Universidad Aut\'onoma de 
Madrid, 28049 Madrid, Spain\label{uam}
\and
Instituto de F\'isica Te\'orica, IFT-UAM/CSIC, C/ Nicolás 
Cabrera 13-15, Universidad Autónoma de Madrid, Cantoblanco, Madrid 
28049, Spain\label{ift}
\and
SISSA -  International School for Advanced Studies, Via Bonomea 265, 34136 Trieste, 
Italy\label{sissa}
\and
Instituto de Astrof\'isica de La Plata (CONICET -- UNLP), Observatorio 
Astron\'omico, Paseo del Bosque S/N, B1900FWA, La Plata, Argentina\label{ialp}
\and
Facultad de Ciencias Astron\'omicas y Geof\'isicas, Universidad 
Nacional de La Plata, Observatorio Astron\'omico, Paseo del Bosque S/N, 
B1900FWA, La Plata, Argentina\label{fcaglp}
\and
Facultad de Ingeniería y Arquitectura, Universidad Central de Chile, 
Av. Francisco de Aguirre 0405, La Serena, Chile.\label{ucc}
}

   \date{Received XXXX, 2024; accepted XXXX, XXXX}

 
  \abstract
   {
   Studies of galaxy populations classified according to their kinematic behaviours and dynamical state using the projected phase space 
   diagram (PPSD) are affected by misclassification and contamination, leading to systematic errors in determining the characteristics of the different galaxy classes.
   }
   {We propose a method for statistically correcting the determination of galaxy 
   properties' distributions that accounts for the contamination caused by misclassified galaxies from other classes.
   }
   {Using a sample of massive clusters and the galaxies in their surroundings taken from the
   \textsc{MultiDark Planck 2} simulation 
   combined with
   the semi-analytic model of galaxy formation \sag, 
   we computed the confusion matrix associated with a classification scheme in the PPSD.
   Based on positions in the PPSD, galaxies are classified as cluster members, backsplash galaxies, recent infallers, 
   infalling galaxies, or interlopers.
   This classification is determined using probabilities calculated by the code \roger\, along with a threshold criterion.
   By inverting the confusion matrix, we are able
   to get better determinations of distributions of galaxy properties, such as colour.}
   {Compared to a direct estimation based solely on the predicted galaxy classes, our method provides better estimates of the mass-dependent colour distribution for the galaxy classes most affected by misclassification: cluster members, backsplash galaxies, and recent infallers.
   We applied the method to a sample of observed X-ray clusters and galaxies.}
   {Our method can be applied to any classification of galaxies in the PPSD, and to any
   other galaxy property besides colour, provided an estimation of the confusion 
   matrix is available. Blue, low-mass galaxies in clusters are almost exclusively recent infaller galaxies that have not 
   yet been quenched by the environmental action of the cluster. Backsplash galaxies
   are on average redder than expected.}

   \maketitle
%

\section{Introduction}

In recent years, the projected phase space 
   diagram (PPSD) has increasingly been used as a
tool for classifying galaxies according to their kinematic behaviours and dynamical state in and around 
systems of galaxies, such as clusters and groups.
The PPSD is a 2D space 
that combines the projected cluster-centric (or group-centric) distance with the line-of-sight velocity relative to the cluster (or group).
Several authors have utilised the PPSD to study the evolution of galaxies. For example, 
\citet{Mahajan:2011} studied the decrease in the star formation of backsplash galaxies (BSs). 
\citet{Muzzin:2014} focused on the quenching of galaxies at $z\sim 1$. \citet{Muriel:2014} 
investigated the properties of high- and low-velocity galaxies in the outskirts of galaxy 
clusters. \citet{HF:2014} and \citet{Jaffe:2015} studied the gas fraction and ram pressure 
of galaxies in clusters. 
\citet{Oman:2016} studied the relationship between the star formation rates of 
a sample of observed galaxies and their likely orbital histories.
\citet{Smith:2019} analysed the stellar mass growth histories of
galaxies as a function of their infall time. \citet{Martinez:2023} studied the quenching of 
galaxies in a sample of X-ray clusters. \citet{Aguerri:2023} focused on the properties of 
barred galaxies and their environments.  \citet{Sampaio:2024} examined the evolution of blue 
cloud, green valley, and red-sequence fractions as a function of time since infall. 
\citet{Munoz:2024} studied the impact on black hole growth as a function of the environment.

The aforementioned studies classify galaxies in systems and their outskirts following 
methods such as those proposed by \citet{Rhee:2017}, \citet{Pasquali:2019}, or 
\citet{delosRios:2021}.
Based on cosmological hydrodynamic {\em N}-body simulations of groups and clusters 
\citep{Choi:2017}, \citet{Rhee:2017} categorised galaxies within the PPSD based on 
the time since their infall into the system. This categorisation included galaxies 
identified as first, recent, intermediate, and ancient infallers. They delineated 
specific regions within the PPSD where each of these types of galaxies are preferentially 
located. 
\citet{Pasquali:2019} present an alternative approach that uses the same cosmological 
simulations \citep{Choi:2017} of groups and clusters employed by \citet{Rhee:2017}. They 
established eight zones of constant mean infall time to investigate environmental influences 
on satellite galaxies. These zones are defined by analytical curves that characterise the 
timing of galaxy infall into the system. 

Recently, \citet{delosRios:2021} presented the code \roger\footnote{\url{https://github.com/Martindelosrios/ROGER}}, which employs three 
different machine learning techniques to classify galaxies based on their projected phase-space positions within and around clusters. They utilised a dataset comprising massive and 
isolated galaxy clusters from the \textsc{MultiDark Planck} 2 simulation 
\citep{klypin_mdpl2_2016}. 
This code establishes a connection between the 2D phase space position of 
galaxies and their 3D orbital classifications. \roger\ calculates the 
probabilities of galaxies belonging to any of the five orbital classes: cluster members 
(CLs), BSs, recent infallers (RINs), infallers (INs), and interlopers 
(ITLs). 

Regardless of the chosen method for classifying galaxies in the PPSD, there will 
inevitably be some degree of contamination among the classes, particularly in regions 
closer to the centres of clusters. According to \citet{Coenda:2022}, misclassifications 
have significant implications for interpreting the properties of galaxies in the 2D 
classes. Therefore, the degree of contamination significantly impacts the conclusions that 
can be drawn. They emphasise the importance of distinguishing between red and blue galaxy 
populations to achieve a more precise analysis of observational data. Furthermore, they 
argue that the 2D analysis reliably provides results only for RINs, INs, and ITLs within the blue population.

To improve determinations of distributions of galaxy observables, such as colours or the
specific star formation rate, for galaxies that have been classified using their
position in the PPSD, we propose a simple method 
that reconstructs the intrinsic distribution of galaxy properties.
The method relies on a statistic estimation of how well the classification is performed.
This paper is organised as follows: In Sect. \ref{sec:method} we present our method; 
in Sect. \ref{sec:sag} we test the method on 
a sample of simulated clusters and galaxies; in Sect. \ref{sec:Xrays} we use the method
to reconstruct the distributions of colours of galaxies of different classes in and around
a sample of X-ray clusters of galaxies; finally, we present our conclusions and mention
possible further uses of the method in Sect. \ref{sec:conclu}.

\section{Method}
\label{sec:method}

If we consider $n$ classes of galaxies according to the particular classification scheme
based on the PPSD we are using, the intrinsic class of each galaxy cannot be determined based solely on its position on it.  
At best, we can assign a particular class 
to a galaxy based on its position in the PPSD using our chosen methodology. In our case, this involves utilising
the output probabilities of
\roger\ combined with a numerical criterion on their values, as we explain in Sect. \ref{sec:sag}.
This way, each galaxy is assigned a predicted class. Due to uncertainties inherent 
to the classification procedure, the number of galaxies in the $i-$th predicted class
($i=1,\ldots,n$), $N^\mathrm{P}_i$, is built up from the contribution of galaxies, in
principle from all intrinsic classes, that were classified as class $i$:
\begin{equation}
    N^\mathrm{P}_i=\sum_{j=1}^{n}C_{ij}N^\mathrm{I}_j,
\end{equation}
where $N^\mathrm{I}_j$ is the number of galaxies in our sample that belong to 
the $j-$th intrinsic class, and $C_{ij}$ is the fraction of galaxies of the $j-$th 
intrinsic class that were classified as being part of the $i-$th predicted class.
The ideal case is $C_{ij}=\delta_{ij}$, that is, a Kronecker delta.
The quantities $C_{ij}$ are the elements of the confusion matrix, 
$\mathbf{C}$, which has dimensions 
$n\times n$.

The primary drawback of imperfect classification arises when studying the characteristics of galaxy populations using predicted classes, a method extensively used in the literature.
Without any knowledge of $\mathbf{C}$, results from
such studies can be misleading. If we want to determine the
distribution of a particular galaxy property, $x$ (e.g. colour, specific star formation rate,
size, etc) for galaxies of a given stellar mass or, alternatively, absolute magnitude ($M$), the observed distributions of the property, $x$, are obtained 
by counting galaxies of a predicted class in $m$ bins of $x$ at a fixed $M$. The fraction of
galaxies of the $i-$th predicted class with $x_k-\Delta x_k/2\leq x \leq x_k+\Delta x_k/2$ 
and normalised to have unity area when integrating over the whole range of $x$ is
\begin{equation}
    f_{ik}(M) =\sum_{j=1}^{n}C_{ij} F_{jk}(M), 
    \label{eq1}
\end{equation}
where the intrinsic (desired) quantities are the $F_{jk}(M)$.
The normalisation condition is\begin{equation}
    \sum_{k=1}^{m} f_{ik}(M) \Delta x_k=1.
\end{equation}
There is no need for the bins in $x$ to have equal size as long as we use the same binning
for all classes.

If we have an estimation of $\mathbf{C}$, for instance, by using numerical simulations,
we can use Eq. \ref{eq1} to solve for $F_{jk}(M)$. By arranging the quantities $f_{ik}(M)$ in a 
$n\times m$ matrix, $\mathbf{f}$, with rows $i$ (classes) and columns $k$ (bins in $x$), and doing the same
for $F_{jk}(M)$ in $\mathbf{F}$, now Eq. \ref{eq1} reads
\begin{equation}
    \mathbf{f}(M) =\mathbf{C}\times \mathbf{F}(M).
    \label{eq2}
\end{equation}
Assuming $\det(\mathbf{C})\neq 0$, the desired distributions are obtained as the rows
of the $n\times m$ matrix:
\begin{equation}
    \mathbf{F}(M) =\mathbf{C}^{-1}\times \mathbf{f}(M).
    \label{eq3}
\end{equation}

It is important to remark that so far we have assumed that the confusion matrix does not 
depend on mass (or absolute magnitude). In Sect. \ref{sec:C(M)} we explore whether a
stellar mass dependence of the confusion matrix (i.e. $\mathbf{C}\equiv \mathbf{C}(M)$) 
provides better results.

In the next section, we test the reliability of our method on a sample of simulated 
massive clusters of galaxies. 

\section{Application to a simulated sample of clusters of galaxies}
\label{sec:sag}

\subsection{The sample of simulated clusters and galaxies}

We used the same sample of clusters and galaxies used by
\citet{delosRios:2021} to train and test the code \roger.
The sample of cluster includes 34 massive systems, $M_{200}\geq 10^{15} h^{-1}M_{\odot}$,
from the \textsc{MultiDark Planck 2} simulation \citep{klypin_mdpl2_2016}. 
These clusters were selected to ensure that they do not have a companion halo more massive than 10 
per cent of $M_{200}$ within $5\times R_{200}$. 
The mass $M_{200}$ is estimated by considering an homogeneous mass distribution enclosed by a characteristic radius of a dark matter halo, $R_{200}$, at which the density is equal to $200$ times the critical density of the Universe at the redshift of the system.

The sample of galaxies in these clusters, and in their surroundings, were modelled by means of the
semi-analytic model of galaxy formation \sag\  \citep{cora_sag_2018}. The sample of simulated galaxies
used in \citet{delosRios:2021} includes all \sagmdpl\ galaxies at $z=0$
with stellar mass greater than $10^{8.5} h^{-1}M_{\odot}$ located within
cylindrical volumes centred in the clusters with radius $3\times R_{200}$, and
an extension along the simulated box's $z-$axis of $\pm 3\sigma$ in velocity (Hubble's
plus peculiar), where $\sigma$ is the line-of-sight velocity dispersion of the galaxies in the cluster 
alongside the box's $z-$axis.

Galaxies within these volumes were classified into five intrinsic classes
according to their past orbits around the clusters  \citep[see][for more details]{delosRios:2021}:
\begin{enumerate}
    \item Cluster members: They may have crossed $R_{200}$ several times in the past 
    and are now found orbiting around the cluster centre, with the majority of them
    within $R_{200}$ of the centre.
    \item Backsplash galaxies: Galaxies that have crossed twice $R_{200}$.
    The first time diving in, and the second time on their way out of the cluster, 
    where they are now. Most of them will fall back into the cluster in the future. 
    \item Recent infallers: Galaxies that have crossed $R_{200}$ only once 
    as they dived into the cluster in the past 2 Gyr. Many of them may be BS in the future.
    \item Infalling galaxies: Galaxies that have always been at distances greater
    than $R_{200}$ from the cluster centre, and have negative radial velocity relative to the cluster. 
    These galaxies are falling to the cluster.
    \item Interlopers: Galaxies that have never gotten closer than $R_{200}$ from the cluster centre 
    but, unlike INs, have positive radial velocities relative to the cluster, that is, 
    they are receding
    away from the cluster at $z=0$. We consider these galaxies as objects that will not
    fall into the cluster. They are unrelated to the cluster 
    but can be confused with classes 1--4 
    above in the PPSD (i.e. they are contamination).
\end{enumerate}

Of the total of 30,289 galaxies,
we used half of them to re-train \roger\ (the training set), and the other
half is the test set, which is the subsample of galaxies we used to compute the
confusion matrix $\mathbf{C}$. In \citet{delosRios:2021}, we used $\sim 80\%$
of the complete sample as training set, and the remaining as the test set.
Here, we are interested in having a larger test set; thus, for the sake of consistency, 
we re-trained \roger\ accordingly.
It is noteworthy that we used a \textsc{python} version of \roger, 
which we dub \pyroger, that is publicly available at \roger's website.

For each galaxy in the test set, we used \pyroger\ to compute its probability
of being a member of each class, $p_i$ where $i=1,\ldots,5$, and the correspondence between index and class
is as listed above: $1 \leftrightarrow$ CL, $2 \leftrightarrow$ BS, $3 \leftrightarrow$ RIN, 
$4 \leftrightarrow$ IN, and $5 \leftrightarrow$ ITL. \roger\ computes these probabilities using three different
techniques: K-nearest neighbours, support vector machine, and random forest. We refer the reader 
to \citet{delosRios:2021} for details. For the purposes of this work, we used the probabilities that
\pyroger\ computes using the K-nearest neighbours technique.

We used the classification scheme adopted by \citet{Coenda:2022}: the predicted class is defined by 
the highest value of $p_i$, and this value has to be higher than a class-dependent threshold. 
These thresholds are 0.4, 0.48, 0.37, 0.54, and 0.15, for CLs,  BSs, RINs, INs, 
and ITLs, respectively. This scheme is an appropriate balance between sensitivity and precision.

\begin{figure}
\centering
\includegraphics[width= 88mm]{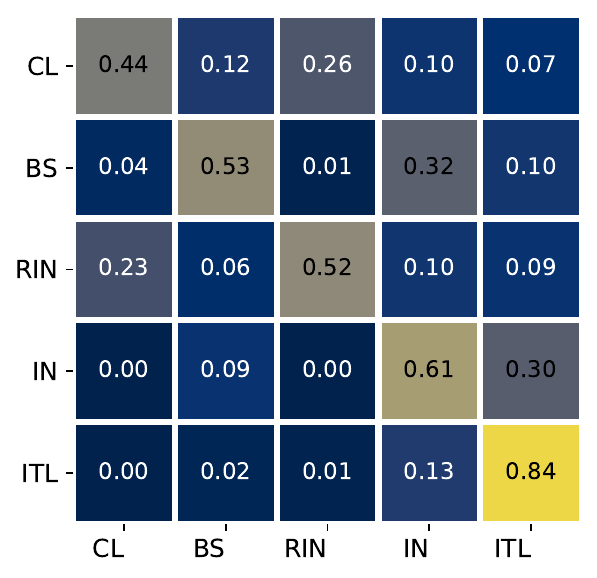}
\caption{Confusion matrix for our adopted classification scheme \citep{Coenda:2022}. 
Columns correspond to intrinsic classes, rows to predicted classes.}
\label{fig:matrix}
\end{figure}

\begin{figure*}
\centering
\includegraphics[width=18cm]{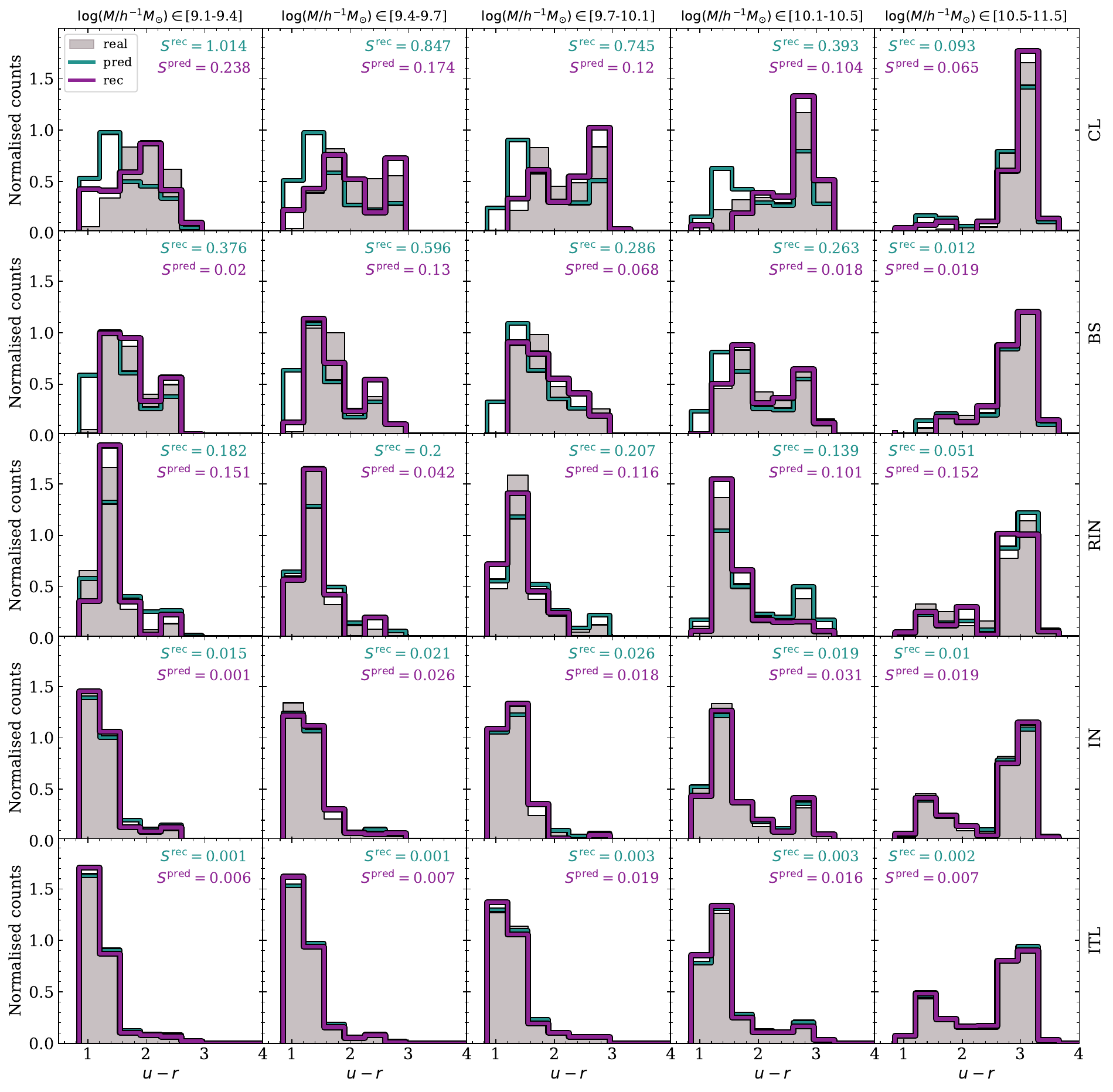}
\caption{Colour distributions of galaxies. Each class is shown in a different row, as noted to the right of the panels. Each column considers a particular range of galaxy stellar mass,
as noted at the top. Grey shaded histograms are the real colour distributions, 
i.e. they correspond to the intrinsic classes. Green lines are the colour distributions of the
predicted classes. Violet lines are the colour distributions recovered by our method. 
We quote within each panel the values
of the corresponding square residuals, $S^{\rm pred}$ and $S^{\rm rec}$, given by
Eqs. \ref{eq:Spred} and \ref{eq:Srec}, respectively.
}
\label{fig:inv_sag_th}
\end{figure*}

\begin{figure}
\centering
\includegraphics[width=6cm]{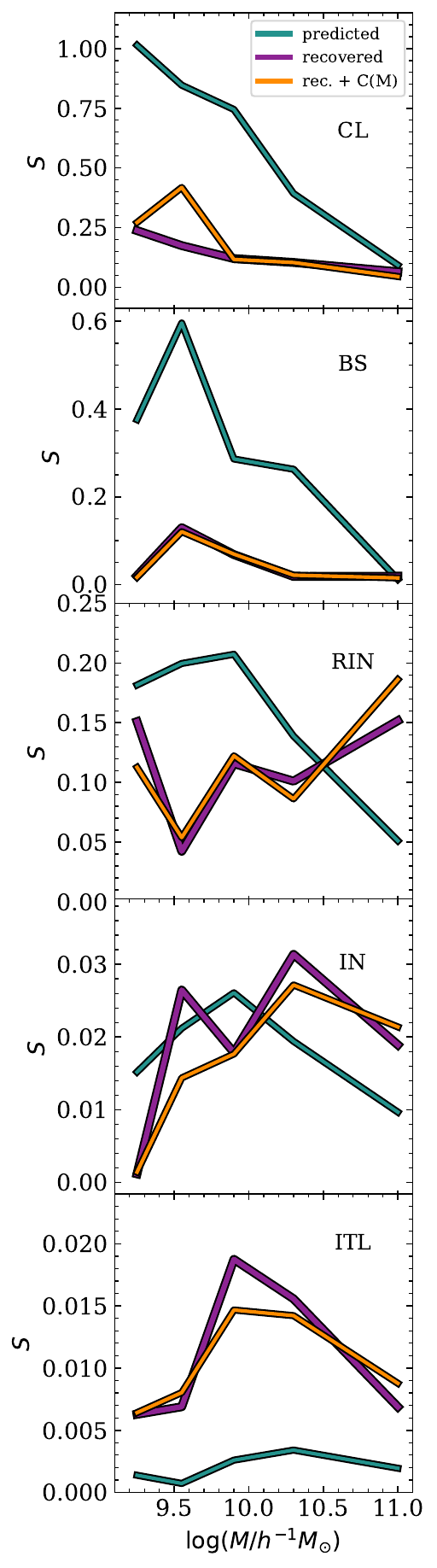}
\caption{Sum of square residuals as a function of galaxy stellar mass. 
Each panel considers a different galaxy class.
Residuals between the real and the predicted distributions are shown with green 
lines, i.e. $S^{\rm pred}$ in Eq. \ref{eq:Spred}. 
Residuals between the real and the recovered distributions are shown with violet lines,
i.e. $S^{\rm rec}$ in Eq. \ref{eq:Srec}. 
The residuals between the real and the recovered
distributions when we use a stellar mass-dependent confusion matrix are shown in orange.
Note that the $y-$axis scale varies from panel to panel.
}
\label{fig:inv_sag_S}
\end{figure}

\subsection{Applying the method}
\label{sec:SAG_inversion}

As a first step, we constructed the confusion matrix using the test set and computed $C_{ij}$,
the fraction of galaxies of predicted class $i$ that are intrinsically of class $j$.
For a better visualisation, instead of simply providing the numbers, we show the confusion matrix
in Fig. \ref{fig:matrix}. 
We can see in Fig. \ref{fig:matrix} how the predicted classes are composed in terms of the real classes.
Predicted ITLs are the simplest case; they are basically real ITLs with a mild contamination
($13\%$) of INs. Predicted INs are roughly real INs with two sources of contamination: $30\%$ ITLs
and 9\% BSs. As for the predicted RINs, 52\% are real RINs, with an unsurprising main source of contamination
from CLs (23\%) and lower ($\leq 10\%$) percentages from the other classes. 
Of the predicted BSs, $53\%$ are real BSs, and their main contamination comes from misclassified INs ($32\%$).
Finally, only $44\%$ of the predicted CLs are real CLs; their most significant contamination is from RINs ($26\%$),
and in second place BSs and INs contributing $\sim 10\%$ each. 
We refer the reader to \citet{delosRios:2021} for a thorough discussion on
how the contamination between the five classes works.
The three most contaminated predicted
classes, CLs, BSs, and INs, are the most interesting ones for studying the effects of environments
in the evolution of galaxies.

In Fig. \ref{fig:inv_sag_th} we present the distributions of $x=u-r$ colour for the intrinsic classes, the predicted classes, and the reconstructed distributions obtained through the inversion of the confusion matrix, according to the adopted classification scheme.
We split galaxies into five stellar mass bins in
our analysis. To quantify whether our method provides
better estimations of the intrinsic colour distribution, we computed 
the sum of squared residuals for each class $i$. This was done by comparing the intrinsic distributions ($F_{ik}(M)$, where $k=1,\ldots,m$ denote the bins in colour) with the colour distributions of both the predicted classes ($f_{ik}(M)$) and the distributions recovered by our method, $F_{ik}^{\rm rec}(M)$:
\begin{equation}
    S_i^{\rm pred}(M)=\sum_{k=1}^m\Big[f_{ik}(M)-F_{ik}(M)\Big]^2,
    \label{eq:Spred}
\end{equation}
and
\begin{equation}
    S_i^{\rm rec}(M)=\sum_{k=1}^m\Big[F_{ik}^{\rm rec}(M)-F_{ik}(M)\Big]^2.
    \label{eq:Srec}
\end{equation}
We note that we computed the square residuals and not $\chi^2$ since there are bins where $F_{ik}(M)=0$. 
These quantities are quoted inside each panel in Fig. \ref{fig:inv_sag_th}. For a better
visualisation and comparison, we show in Fig. \ref{fig:inv_sag_S} the sum of square residuals as a function
of stellar mass. We note that $y-$axis scale varies from panel to panel in Fig. \ref{fig:inv_sag_S}.

A joint analysis of Figs. \ref{fig:inv_sag_th} and \ref{fig:inv_sag_S} gives two general conclusions:
on the one hand, our method improves the determination of the colour distribution of the three predicted classes
that are more affected by misclassifications, namely, CLs, BSs, and RINs; and on the other hand,
it does not improve the results for INs and ITLs, which are the two predicted classes best classified
by our adopted criteria. 
These two main results are promising for this type of study; however, there is more to consider beyond just two levels of contamination, namely high contamination for CLs, BSs, and RINs and low contamination for INs and ITLs. 
Regarding their intrinsic colour distributions,
the IN and ITL classes are the most similar,
while the other three classes show greater variations
from one to another. This is the consequence of very different evolutionary paths.
Delving into the specifics, there are several interesting details in Figs. \ref{fig:inv_sag_th}
and \ref{fig:inv_sag_S}, which we summarise below.
\begin{enumerate}
    \item Cluster members: Our method improves greatly the determination of the colour distributions in all five stellar
    mass bins. The predicted CL class exhibits a contamination from blue galaxies, which is particularly strong
    in the lower-mass bins. These galaxies are mostly RINs, and in a second place BSs, incorrectly classified as
    CLs. Distributions recovered by our method eliminate most of this contamination. We consistently find 
    $S_{\rm CL}^{\rm rec}(M)<S_{\rm CL}^{\rm pred}(M)$ for all mass bins.  
    \item Backsplash galaxies: As in the previous case, there are clear improvements for BS galaxies too, consisting in the 
    removal of the contamination from blue, and mainly INs, galaxies. For all mass bins we find 
    $S_{\rm BS}^{\rm rec}(M)<S_{\rm BS}^{\rm pred}(M)$. Backsplash galaxies are, on average, redder when contamination effects are removed by our method.
    \item Recent infallers: With the exception of the highest-mass bin, we find that the method produces 
    better results (i.e.
    $S_{\rm RIN}^{\rm rec}(M)<S_{\rm RIN}^{\rm pred}(M))$. The improvement is most notably seen in the removal 
    of the contamination from misclassified red CL galaxies with stellar mass in the
    range $9.7\leq \log(M/h^{-1} M_{\odot})\leq 10.5$ (see the third and 
    fourth columns in Fig. \ref{fig:inv_sag_th}). Our findings for CLs 
    above and for RINs here suggest that almost every blue galaxy found in a cluster should have dived in
    there recently. This could have implications on the characterisation of satellites in clusters according to 
    their dynamical state (e.g. \citealt{Aldas:2023,Aldas:2024}).
    \item Infalling galaxies: In this case the method gives results comparable to the raw predictions. We can observe
    no significant differences between the intrinsic, the predicted, and the recovered distributions in Fig.
    \ref{fig:inv_sag_S}. The values of the sum of square residuals are 
    $S_{\rm IN}^{\rm rec}(M)>S_{\rm IN}^{\rm pred}(M)$ for three alternate mass bins, and otherwise in the
    remaining two. We note that $S_{\rm IN}^{\rm rec}(M)$ and $S_{\rm BS}^{\rm pred}(M)$ are very similar
    in all cases and take values $\sim 1-2 \times 10^{-2}$, which is, on average, smaller than the
    corresponding values we find for CLs, BSs, and RINs.
    \item Interlopers: As said before, for this class, our method provides no improvement at all, but this should
    not be considered as a problem since there are
    virtually no differences
    between the intrinsic, predicted and recovered colour distributions,
    as in the previous case.
    In all five mass bins we obtain
    $S_{\rm ITL}^{\rm rec}(M)>S_{\rm ITL}^{\rm pred}(M)$, but we note that all numerical values ($\sim 10^{-3}$) are an 
    order of magnitude smaller than for INs.
\end{enumerate}

\subsection{The effects of a mass dependence of matrix $\mathbf{C}$}
\label{sec:C(M)}

We explored the effect on our method of a possible dependence of the confusion 
matrix on the mass of galaxies. For each mass bin considered in Fig. \ref{fig:inv_sag_th}, we computed a
confusion matrix, $\mathbf{C}(M)$, using only those galaxies in the test sample that have stellar mass
in the bin. The matrices so obtained (not shown here) do have a mild dependence on mass. However, this 
dependence do not change the results  discussed
in the previous subsection in a significant way.
In Fig. \ref{fig:inv_sag_S}, we present the resulting values of $S_i^{\rm rec}(M)$ as orange lines, where it is evident that the differences compared to the case of a unique $\mathbf{C}$ are really small.
We can safely affirm there is no need to use a mass-dependent (or absolute-magnitude-dependent) confusion matrix. Furthermore, a
mass-dependent confusion matrix implies a greater dependence on the specifics of how the
galaxy evolution model works. It is desirable to minimise reliance on the model as much as possible.

The confusion matrix may depend on multiple parameters involved in the classification procedure, not just the stellar mass of the galaxies. In our particular case, classification was performed using specific thresholds on the probabilities computed by our code \roger, which was trained on massive clusters from a specific cosmological simulation combined with a particular semi-analytic model.
It is clear that the confusion matrix depends not only on these thresholds but also on the choices made during the training process. Given the nature of the input data used for classification, namely, the position in the PPSD, we do not expect the specific details of the semi-analytic model to have a significant impact on the confusion matrix. However, the choice of the mass range of the clusters used to train \roger\ is likely to have a considerable effect.
Other factors may also play a role, such as the degree of relaxation of the clusters. Assessing the influence of these various factors on the confusion matrix would require a comprehensive study that is beyond the scope of this paper. Nevertheless, the key aspect of the method proposed in this paper lies in constructing a confusion matrix that best represents the specifics of the classification performed. Achieving this will most likely require the use of numerical simulations of the galaxy systems under study.


\section{Application to an observed sample of X-ray clusters of galaxies}
\label{sec:Xrays}

As an example of the potential of the proposed statistical technique to correct the effects of misclassification, 
we applied our method to the sample of X-ray clusters and Sloan Digital Sky Survey (SDSS) galaxies used in
\citet{Martinez:2023}. The sample comprises 104 X-ray clusters drawn from the works by \citet{Coenda:2009}
and \citet{Muriel:2014}, where the authors compute a number of properties of 
clusters originally catalogued by \citet{Popesso:2004} and \citet{Noras:2000}, 
respectively. \citet{Martinez:2023} searched for all spectroscopic SDSS-DR7 \citep{dr7} galaxies
within cylindrical volumes in redshift space centred in the clusters, and with radius $3\times R_{200}$
in projection and a longitude of $\pm 3 \times \sigma$ alongside the line-of-sight of the cluster.
Galaxies were classified in the five classes under study using their positions in the PPSD, 
the \roger-computed probabilities, and the thresholds of \citet{Coenda:2022}. 

Using the confusion matrix computed in Sect. \ref{sec:SAG_inversion},
we estimated the predicted and recovered distributions of the
$^{0.1}(u-r)$ colour of the five classes in four bins in stellar mass. The resulting distributions
are shown in Fig. \ref{fig:inv_xrays}. The outcome of applying our method to this sample of
observed galaxies and clusters qualitatively agrees with our findings in Sect. 
\ref{sec:SAG_inversion}, which were derived from applying the method to a simulated galaxy catalogue. We can observe basically the same trends of Fig. \ref{fig:inv_sag_S}.
For CL, the distributions recovered by our method eliminate most of the contribution 
from blue galaxies, this is more notorious in the lower-mass bins. According to the insights of Sect. 
\ref{sec:SAG_inversion}, these blue galaxies are mostly actual RINs. In the case of BSs, the main feature we obtain is the removal of the contamination from blue galaxies, particularly at the 
lowest-mass bin. According to the confusion matrix we used (see Sect. \ref{sec:SAG_inversion}), 
some $70\%$ of the contamination should be due to galaxies erroneously classified as INs. 
Regarding RINs, the result of our method is most notably seen in the removal 
of a population of red, almost exclusively CLs, most clearly in the lower-mass bins.  
As in Sect. \ref{sec:SAG_inversion}, for INs, the method gives 
results very similar to the raw predictions. Finally, for ITLs, once again, 
there are virtually no differences between the predicted and recovered colour distributions.


\section{Conclusions}
\label{sec:conclu}

We propose a method for recovering the distributions of observed properties of galaxies classified according
to their position in the PPSD. The method requires knowledge of the
confusion matrix, or at least an estimation of it, which is inverted and applied to observed
distributions.
We tested this method on simulated clusters and galaxies and applied it to a sample of galaxies and X-ray 
clusters from the SDSS.
For both simulated and observed galaxies, we classified galaxies in the PPSD using the 
code \roger\ from \citet{delosRios:2021} plus the criteria from \citet{Coenda:2022}.
In particular, we utilised colour as the parameter for analysing the distributions. These two 
choices, namely (i) the classification using \roger\ plus thresholds and (ii) colour, are not essential to our 
objectives.
Any classification scheme over the PPSD, such as those by \citet{Rhee:2017} and 
\citet{Pasquali:2019}, and any galaxy property would be suitable provided one has 
an estimation of the confusion matrix.

Using the inversion of the confusion matrix along with our \roger-based classification provides
a better estimation of the distribution of colours of galaxies in and around clusters. The method
is successful at decontaminating different classes of galaxies from the contamination of other classes,
and it is particularly useful for regions of the PPSD where the overlap between classes
complicates the classification most. 
The proposed method enables further analyses of galaxies when the classification is an issue.
It is straightforward to reconstruct some statistics of galaxy populations as a function of stellar
mass (or absolute magnitude). For instance, from Fig. \ref{fig:inv_xrays}, we can
compute the median of the colour distribution as a function of mass for the blue and red galaxy populations
of each galaxy class. 

We can draw some conclusions from the results we obtain for the five classes.
The bulk of galaxies in clusters that are blue and less massive than $M \simeq 10^{10.5} h^{-1}M_{\odot}$ 
are RINs.
In terms of colour, BSs are on average in between CLs and RINs, while being distinctly redder than INs.
Thus, a single excursion through the inner
regions of a cluster produces significant changes in galaxy colour, due to a quench in star formation 
\citep[e.g.][]{Hough:2023,Ruiz:2023}. 

It is crucial to emphasise that analysing the properties of galaxies in systems derived directly from orbits obtained through phase space can lead to erroneous conclusions due to contamination effects between different classes. Therefore, the implementation of techniques aimed at minimising these biases, such as the one presented in this work, is highly important.

\begin{acknowledgements}
This paper has been partially supported with grants from Consejo Nacional de
Investigaciones Cient\'ificas y T\'ecnicas (PIPs 11220130100365CO,
11220210100064CO, 11220200102832CO and 11220200102876CO), Argentina, the Agencia Nacional de
Promoci\'on Cient\'ifica y Tecnol\'ogica (PICTs 2018-3743, 2020-3690 and 2021-I-A-00700),
Argentina, Secretar\'ia de Ciencia y Tecnolog\'ia, Universidad Nacional de
C\'ordoba, Argentina, and Universidad Nacional de La Plata (G11-183), Argentina.
MdlR acknowledges financial support from the Comunidad Aut\'onoma de
Madrid through the grant SI2/PBG/2020-00005. 
MdlR is supported by
the Next Generation EU program, in the context of the National Recovery and 
Resilience Plan, Investment PE1 – Project FAIR “Future Artificial Intelligence Research”.
The \textsc{CosmoSim} database used in this paper is a service by the
Leibniz-Institute for Astrophysics Potsdam (AIP). The \textsc{MultiDark}
database was developed in cooperation with the Spanish MultiDark Consolider
Project CSD2009-00064.  The authors gratefully acknowledge the Gauss Centre for
Supercomputing e.V. (www.gauss-centre.eu) and the Partnership for Advanced
Supercomputing in Europe (PRACE, www.prace-ri.eu) for funding the
\textsc{MultiDark} simulation project by providing computing time on the GCS
Supercomputer SuperMUC at Leibniz Supercomputing Centre (www.lrz.de). 
\end{acknowledgements}

\bibliographystyle{aa} 

\begin{thebibliography}{26}
\expandafter\ifx\csname natexlab\endcsname\relax\def\natexlab#1{#1}\fi

\bibitem[{{Abazajian} {et~al.}(2009){Abazajian}, {Adelman-McCarthy},
  {Ag{\"u}eros}, {Allam}, {Allende Prieto}, {An}, {Anderson}, {Anderson},
  {Annis}, {Bahcall}, \& et~al.}]{dr7}
{Abazajian}, K.~N., {Adelman-McCarthy}, J.~K., {Ag{\"u}eros}, M.~A., {et~al.}
  2009, \apjs, 182, 543

\bibitem[{{Aguerri} {et~al.}(2023){Aguerri}, {Cuomo}, {Rojas-Roncero}, \&
  {Morelli}}]{Aguerri:2023}
{Aguerri}, J. A.~L., {Cuomo}, V., {Rojas-Roncero}, A., \& {Morelli}, L. 2023,
  \aap, 679, A5

\bibitem[{{Ald{\'a}s} {et~al.}(2024){Ald{\'a}s}, {G{\'o}mez},
  {Vega-Mart{\'\i}nez}, {Zenteno}, \& {Carrasco}}]{Aldas:2024}
{Ald{\'a}s}, F., {G{\'o}mez}, F.~A., {Vega-Mart{\'\i}nez}, C., {Zenteno}, A.,
  \& {Carrasco}, E.~R. 2024, arXiv e-prints, arXiv:2408.05305

\bibitem[{{Ald{\'a}s} {et~al.}(2023){Ald{\'a}s}, {Zenteno}, {G{\'o}mez},
  {Hernandez-Lang}, {Carrasco}, {Vega-Mart{\'\i}nez}, \& {Nilo
  Castell{\'o}n}}]{Aldas:2023}
{Ald{\'a}s}, F., {Zenteno}, A., {G{\'o}mez}, F.~A., {et~al.} 2023, \mnras, 525,
  1769

\bibitem[{{Bohringer} {et~al.}(2000){Bohringer}, {Voges}, {Huchra}, {McLean},
  {Giacconi}, {Rosati}, {Burg}, {Mader}, {Schuecker}, {Simic}, {Komossa},
  {Reiprich}, {Retzlaff}, \& {Trumper}}]{Noras:2000}
{Bohringer}, H., {Voges}, W., {Huchra}, J.~P., {et~al.} 2000, VizieR Online
  Data Catalog, 212, 90435

\bibitem[{{Choi} \& {Yi}(2017)}]{Choi:2017}
{Choi}, H. \& {Yi}, S.~K. 2017, \apj, 837, 68

\bibitem[{{Coenda} {et~al.}(2022){Coenda}, {de los Rios}, {Muriel}, {Cora},
  {Mart{\'\i}nez}, {Ruiz}, \& {Vega-Mart{\'\i}nez}}]{Coenda:2022}
{Coenda}, V., {de los Rios}, M., {Muriel}, H., {et~al.} 2022, \mnras, 510, 1934

\bibitem[{{Coenda} \& {Muriel}(2009)}]{Coenda:2009}
{Coenda}, V. \& {Muriel}, H. 2009, \aap, 504, 347

\bibitem[{{Cora} {et~al.}(2018){Cora}, {Vega-Mart{\'\i}nez}, {Hough}, {Ruiz},
  {Orsi}, {Mu{\~n}oz Arancibia}, {Gargiulo}, {Collacchioni}, {Padilla},
  {Gottl{\"o}ber}, \& {Yepes}}]{cora_sag_2018}
{Cora}, S.~A., {Vega-Mart{\'\i}nez}, C.~A., {Hough}, T., {et~al.} 2018, \mnras,
  479, 2

\bibitem[{{de los Rios} {et~al.}(2021){de los Rios}, {Mart{\'\i}nez}, {Coenda},
  {Muriel}, {Ruiz}, {Vega-Mart{\'\i}nez}, \& {Cora}}]{delosRios:2021}
{de los Rios}, M., {Mart{\'\i}nez}, H.~J., {Coenda}, V., {et~al.} 2021, \mnras,
  500, 1784

\bibitem[{{Hern{\'a}ndez-Fern{\'a}ndez}
  {et~al.}(2014){Hern{\'a}ndez-Fern{\'a}ndez}, {Haines}, {Diaferio},
  {Iglesias-P{\'a}ramo}, {Mendes de Oliveira}, \& {Vilchez}}]{HF:2014}
{Hern{\'a}ndez-Fern{\'a}ndez}, J.~D., {Haines}, C.~P., {Diaferio}, A., {et~al.}
  2014, \mnras, 438, 2186

\bibitem[{{Hough} {et~al.}(2023){Hough}, {Cora}, {Haggar}, {Vega-Martinez},
  {Kuchner}, {Pearce}, {Gray}, {Knebe}, \& {Yepes}}]{Hough:2023}
{Hough}, T., {Cora}, S.~A., {Haggar}, R., {et~al.} 2023, \mnras, 518, 2398

\bibitem[{{Jaff{\'e}} {et~al.}(2015){Jaff{\'e}}, {Smith}, {Candlish},
  {Poggianti}, {Sheen}, \& {Verheijen}}]{Jaffe:2015}
{Jaff{\'e}}, Y.~L., {Smith}, R., {Candlish}, G.~N., {et~al.} 2015, \mnras, 448,
  1715

\bibitem[{{Klypin} {et~al.}(2016){Klypin}, {Yepes}, {Gottl{\"o}ber}, {Prada},
  \& {He{\ss}}}]{klypin_mdpl2_2016}
{Klypin}, A., {Yepes}, G., {Gottl{\"o}ber}, S., {Prada}, F., \& {He{\ss}}, S.
  2016, \mnras, 457, 4340

\bibitem[{{Mahajan} {et~al.}(2011){Mahajan}, {Mamon}, \&
  {Raychaudhury}}]{Mahajan:2011}
{Mahajan}, S., {Mamon}, G.~A., \& {Raychaudhury}, S. 2011, \mnras, 416, 2882

\bibitem[{{Mart{\'\i}nez} {et~al.}(2023){Mart{\'\i}nez}, {Coenda}, {Muriel},
  {de los Rios}, \& {Ruiz}}]{Martinez:2023}
{Mart{\'\i}nez}, H.~J., {Coenda}, V., {Muriel}, H., {de los Rios}, M., \&
  {Ruiz}, A.~N. 2023, \mnras, 519, 4360

\bibitem[{{Mu{\~n}oz Rodr{\'\i}guez} {et~al.}(2024){Mu{\~n}oz Rodr{\'\i}guez},
  {Georgakakis}, {Shankar}, {Ruiz}, {Bonoli}, {Comparat}, {Fu}, {Koulouridis},
  {Lapi}, \& {Almeida}}]{Munoz:2024}
{Mu{\~n}oz Rodr{\'\i}guez}, I., {Georgakakis}, A., {Shankar}, F., {et~al.}
  2024, \mnras, 532, 336

\bibitem[{{Muriel} \& {Coenda}(2014)}]{Muriel:2014}
{Muriel}, H. \& {Coenda}, V. 2014, \aap, 564, A85

\bibitem[{{Muzzin} {et~al.}(2014){Muzzin}, {van der Burg}, {McGee}, {Balogh},
  {Franx}, {Hoekstra}, {Hudson}, {Noble}, {Taranu}, {Webb}, {Wilson}, \&
  {Yee}}]{Muzzin:2014}
{Muzzin}, A., {van der Burg}, R.~F.~J., {McGee}, S.~L., {et~al.} 2014, \apj,
  796, 65

\bibitem[{{Oman} \& {Hudson}(2016)}]{Oman:2016}
{Oman}, K.~A. \& {Hudson}, M.~J. 2016, \mnras, 463, 3083

\bibitem[{{Pasquali} {et~al.}(2019){Pasquali}, {Smith}, {Gallazzi}, {De Lucia},
  {Zibetti}, {Hirschmann}, \& {Yi}}]{Pasquali:2019}
{Pasquali}, A., {Smith}, R., {Gallazzi}, A., {et~al.} 2019, \mnras, 484, 1702

\bibitem[{{Popesso} {et~al.}(2004){Popesso}, {B{\"o}hringer}, {Brinkmann},
  {Voges}, \& {York}}]{Popesso:2004}
{Popesso}, P., {B{\"o}hringer}, H., {Brinkmann}, J., {Voges}, W., \& {York},
  D.~G. 2004, \aap, 423, 449

\bibitem[{{Rhee} {et~al.}(2017){Rhee}, {Smith}, {Choi}, {Yi}, {Jaff{\'e}},
  {Candlish}, \& {S{\'a}nchez-J{\'a}nssen}}]{Rhee:2017}
{Rhee}, J., {Smith}, R., {Choi}, H., {et~al.} 2017, \apj, 843, 128

\bibitem[{{Ruiz} {et~al.}(2023){Ruiz}, {Mart{\'\i}nez}, {Coenda}, {Muriel},
  {Cora}, {de los Rios}, \& {Vega-Mart{\'\i}nez}}]{Ruiz:2023}
{Ruiz}, A.~N., {Mart{\'\i}nez}, H.~J., {Coenda}, V., {et~al.} 2023, \mnras,
  525, 3048

\bibitem[{{Sampaio} {et~al.}(2024){Sampaio}, {de Carvalho},
  {Arag{\'o}n-Salamanca}, {Merrifield}, {Ferreras}, \&
  {Cornwell}}]{Sampaio:2024}
{Sampaio}, V.~M., {de Carvalho}, R.~R., {Arag{\'o}n-Salamanca}, A., {et~al.}
  2024, \mnras, 532, 982

\bibitem[{{Smith} {et~al.}(2019){Smith}, {Pacifici}, {Pasquali}, \&
  {Calder{\'o}n-Castillo}}]{Smith:2019}
{Smith}, R., {Pacifici}, C., {Pasquali}, A., \& {Calder{\'o}n-Castillo}, P.
  2019, \apj, 876, 145

\end{thebibliography}


\clearpage
\begin{appendix}
\section{Additional figure}
\begin{figure}[h]
\centering
\includegraphics[width=0.9\textwidth]{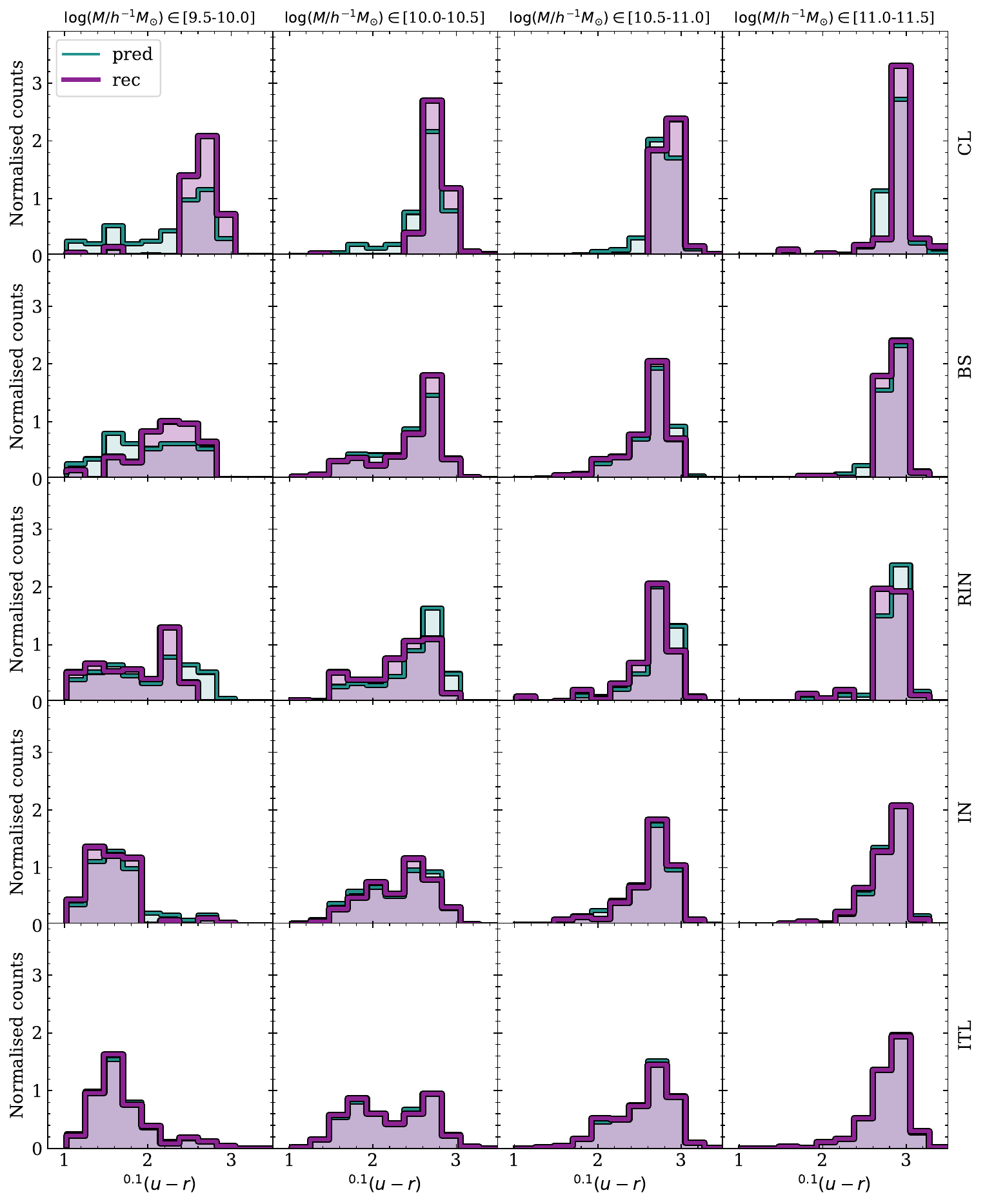}
\caption{
Application of our method to the sample of galaxies in and around X-ray clusters
used by \citet{Martinez:2023}. As in Fig. \ref{fig:inv_sag_th}, rows
show different classes, and columns different stellar mass ranges.
Green shaded histograms are predicted colour distributions and violet shaded histograms recovered colour distributions.
}
\label{fig:inv_xrays}
\end{figure}

\clearpage
\end{appendix}

\end{document}